\documentclass[10pt,conference]{IEEEtran}
\IEEEoverridecommandlockouts
\AtBeginDocument{%
  \providecommand\BibTeX{{%
    \normalfont B\kern-0.5em{\scshape i\kern-0.25em b}\kern-0.8em\TeX}}}

\usepackage{graphicx}
\usepackage{todonotes}
\usepackage{balance}
\usepackage{enumitem}
\usepackage{caption}
\usepackage{url}
\usepackage{booktabs}
\usepackage{titlesec}
\titlespacing*{\subsection}
{0pt}{5pt}{3pt}
\makeatletter
\renewcommand{\paragraph}{%
  \@startsection{paragraph}{4}%
  {\z@}{0.5ex \@plus 0.25ex \@minus .2ex}{-1em}%
  {\normalfont\normalsize\itshape\bfseries}%
}
\makeatother

\begin{document}

\title{ResearchBot: Bridging the Gap between Academic Research and Practical Programming Communities}


\author{
\IEEEauthorblockN{Sahar Farzanepour\textsuperscript{*}, Swetha Rajeev\textsuperscript{*}, Huayu Liang\textsuperscript{*}, Ritvik Prabhu, and Chris Brown}
\IEEEauthorblockA{\textit{Department of Computer Science} \\
\textit{Virginia Tech,}
Blacksburg, VA, USA \\
\{saharfarza, rswetha, huayu98, ritvikp, dcbrown\}@vt.edu}
\thanks{*These authors contributed equally to this work}
}


\maketitle

\begin{abstract}
Software developers commonly rely on platforms like Stack Overflow for problem-solving and learning. However, academic research is an untapped resource that could greatly benefit industry practitioners. The challenge lies in connecting the innovative insights from academia to real-world problems faced by developers. This project introduces \textit{ResearchBot}, a tool designed to bridge this academia-industry gap.
\textit{ResearchBot} employs a modular approach, encompassing understanding questions, curating queries to obtain relevant papers in the CrossRef repository, summarizing paper content, and finally answering user questions based on paper summaries. The core objective of \textit{ResearchBot} is to democratize access to academic knowledge for industry professionals. By providing concise summaries of cutting-edge research directly in response to SE-related questions, \textit{ResearchBot} facilitates the application of academic insights to practical contexts. Ultimately, it aims to bridge the gap between academia and industry, using research evidence to support learning and decision-making in software development.
\end{abstract}







\section{Introduction}
Programmers frequently use Q\&A platforms like Stack Overflow\footnote{\url{https://stackoverflow.com/}} for software engineering support and problem-solving\cite{barua2014developers}. Similarly, research shows developers often rely on Stack Exchange\footnote{\url{https://stackexchange.com/}} to seek information about various software development processes and problems~\cite{sulir2022software}. However, while these resources provide a wide range of user-generated answers, prior work suggests responders in virtual knowledge sharing communities may lack expertise and skills necessary to respond to a wide variety of SE-related questions~\cite{menshikova2018evaluation}.

Recently, large language models (LLMs), e.g., ChatGPT,\footnote{\url{https://chat.openai.com/}} are capable of providing solutions to SE-related tasks due to their ability to provide contextually relevant answers~\cite{leung2023automated,xiao2024devgpt}. In addition, prior work suggests developers prefer responses from ChatGPT over Stack Overflow~\cite{kabir2023answers}. However, these tools typically draw answers from general web sources, including blogs and informal Q\&A sites, rather than scholarly databases. This can limit the reliability and depth of their answers, as ChatGPT’s responses sometimes include broken or irrelevant links and lack connections to academically validated sources. Prior work highlights these limitations in both Stack Overflow~\cite{huang2024tale} and ChatGPT~\cite{auer2023sciqa}, emphasizing the need for novel approaches that align academic research  with practical community needs more effectively. While platforms like Stack Overflow provide quick solutions, ResearchBot offers evidence based insights from scholarly research, ensuring higher reliability and depth for critical decisions in software engineering. 

Academic research can offer distinct advantages for practitioners, such as evidence-based insights to enhance SE~\cite{evidence,arcuri2011practical,kitchenham2004evidence}. However, there is a significant gap between academic research and practical programming communities. Despite the rigorous methods and extensive analyses used in SE research~\cite{storey2020software}, findings from scientific insights often fail to reach practitioners seeking solutions~\cite{rainer2003persuading,devanbu2016belief}. This gap is caused by challenges such as the relevance and credibility of research when applied to the fast-paced field of software development~\cite{winters2024thoughts,lo2015practitioners} in addition to a lack of accessibility due to paywalls, technical language, and the volume of available publications~\cite{wilson2024will,access}. Prior work has explored methods to increase adoption of SE research in practice, such as research summaries of empirical SE findings for practitioners,\footnote{\url{https://neverworkintheory.org/}} ultimately finding these approaches ineffective~\cite{wilson2024will}. Moreover, sources suggest bridging the academia-industry gap is one of the top priorities for the future of AI~\cite{horvitznow}.

To address these challenges, we introduce \textit{ResearchBot}, a chatbot designed to support and answer developer questions based on academic research. \textit{ResearchBot} leverages advanced natural language processing (NLP) techniques to automatically understand user questions, and retrieve relevant papers from the CrossRef repository\footnote{\url{https://www.crossref.org/}}---a comprehensive database containing metadata about scholarly articles, conference proceedings, books, and datasets---and provide concise summaries. This work-in-progress paper presents the results of a preliminary evaluation exploring the capabilities of \textit{ResearchBot} to provide human-like responses to practical SE-related queries. By transforming theoretical knowledge into practical contexts, \textit{ResearchBot} aims to provide a first step in bridging the gap between academia and industry---facilitating the application of research evidence to real-world SE problems~\cite{kitchenham2004evidence}.


\section{Implementation}

\begin{figure*}
    \centering 
    \includegraphics[width=0.70\textwidth]{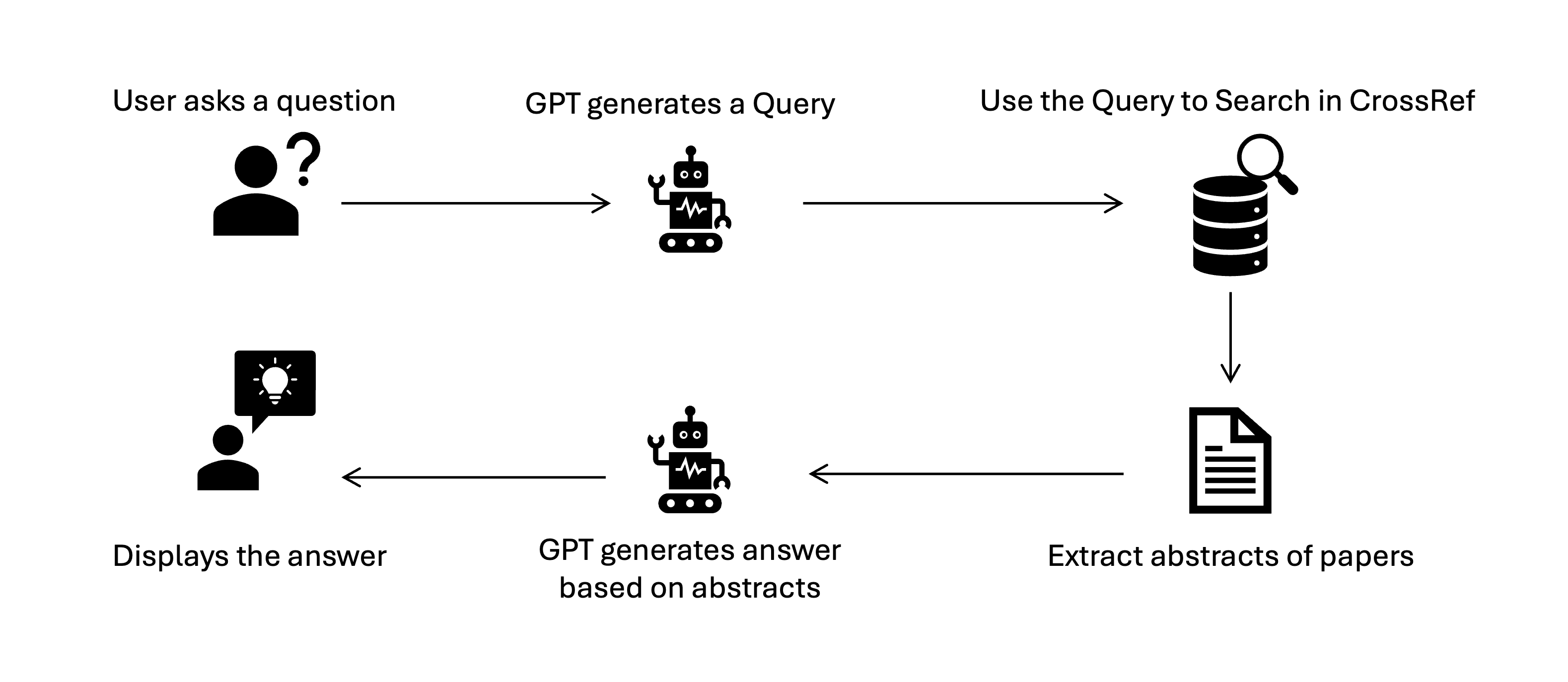} 
    \caption{Pipeline for \textit{ResearchBot}} 
    \label{fig:Pipeline} 
\end{figure*}

The implementation of \textit{ResearchBot} integrates two main modules: the Retrieval Module and the Question-Answer (QA) Module, leveraging GPT-4's capabilities of natural language understanding. This system utilizes specific prompts to guide the model in querying the Crossref repository and generating informative responses based on retrieved research papers. For now, \textit{ResearchBot} is unable to analyze and generate code as it only gives answers to practical questions based on research findings. Figure \ref{fig:Pipeline} illustrates the pipeline of \textit{ResearchBot}, showcasing the flow from query to response generation.

\subsection{Retrieval Module}
The Retrieval Module fetches relevant academic papers based on queries.

Here is the workflow of retrieval module:
\begin{enumerate}[topsep=0pt]
    \item \textit{ResearchBot} receives a user question.
    \item GPT-4 generates a structured query based on the user question, formatted with appropriate keywords for the CrossRef API 
    
    (e.g., ``nlp+natural+language+processing+document+
    
    similarity").
    \item The query is sent to the CrossRef API to retrieve the top 15 most relevant papers.
    \item Abstracts of these papers are extracted for further processing.
\end{enumerate}

Here is the prompt to instruct GPT-4 to query the Crossref API:

\noindent
\texttt{\small From now on I will pass in a prompt like "How do I   
perform document similarity using NLP" and I would 
like you to  curate an output that looks similar to  
this "nlp+natural+language+processing+document+
similarity". This is what a search query would look 
like for the given prompt. It uses `+' which 
separates all the keywords}.

The prompt used in this module guides GPT-4 on how to transform a natural language question into a structured query format. This ensures that the search query generated by the model is specific and optimized for retrieving relevant academic papers. Without a clear prompt, the model might struggle to understand the intended format of the query and could produce less accurate results.

\subsubsection{Paper Abstract Extraction}

To retrieve abstracts for \textit{ResearchBot}, we utilize the CrossRef API\footnote{\url{https://api.crossref.org}} to initialize the query parameters, setting the number of desired results and filtering for entries with abstracts. We then send a request to the API to fetch papers matching the query, sorted by relevance. For a success request, the response data is parsed to extract paper details, including titles, abstracts, URLs, and citation counts. Each abstract is processed using a function (\texttt{\small jats\_to\_plain\_text}) that converts the JATS-formatted XML to plain text to ensure readability. Finally, a list of relevant papers sorted by citation count in descending order is created for further analysis.If no relevant abstracts are found, ResearchBot returns a message indicating insufficient research data. Future iterations will incorporate full-text analysis to improve accuracy.

\subsection{QA Module}
The QA Module of \textit{ResearchBot} processes the retrieved paper abstracts and user question to generate a comprehensive response.

Here is the workflow of QA module:
\begin{enumerate}[topsep=0pt]
    \item The user question and retrieved paper abstracts (along with their URLs) are passed to GPT-4.
    \item GPT-4 generates a concise and informative response, utilizing the context provided by the abstracts to address the user's question.
    \item Citations to relevant papers are embedded within the response text using the specified format.
\end{enumerate}

Here is the prompt to instruct GPT-4 to answer user's question:

\noindent
\texttt{\small From now on I will pass in a Software Engineering  
question and some relevant research papers as  
context for the answer. I would like you to answer 
the question with references made to the contexts   
(make sure that the focus of the answer is to 
answer the question, not the context itself). 
Be sure to also include the URL after the reference 
to the paper in the text. Use this format to cite 
in-text (<a href="url" target="\_blank">title</a>).If your question is outside computer science, ResearchBot cannot provide an answer as it only supports CS-related queries.}

The prompt used in the QA Module provides GPT-4 with clear instructions on how to structure its response. By specifying the focus on answering the user question with reference to the provided context (research paper abstracts), the prompt ensures that the generated response is relevant and informative. LLMs can excel in understanding and generating human-like text, enabling them to process complex queries and produce responses tailored to specific questions~\cite{elastic}. Additionally, the prompt dictates the citation format, enabling proper acknowledgment of the cited sources within the response.



    
    



    

\section{Methodology}

\subsection{Data Collection}
For the purpose of this study, data was systematically collected from the Software Engineering Stack Exchange community.\footnote{\url{https://softwareengineering.stackexchange.com/}} This platform was selected due to its extensive use by programming professionals, who regularly engage in discussions concerning practical challenges and their solutions.
Our selection criteria focused primarily on technical inquiries, particularly those pertaining to subjects such as programming languages and software development. 
We focused on technical inquiries from SE Stack Exchange, avoiding Stack Overflow due to coding-related limitations. Using these criteria, we collected a dataset of 100 question-answer pairs to analyze the functionality of \textit{ResearchBot}.

\urldef\urlSpacy\url{https://spacy.io/usage/linguistic-features#vectors-similarity}
\urldef\urlSBERT\url{https://sbert.net/}

\subsection{Data Analysis}
\subsubsection{Semantic Textual Similarity}
We utilized semantic textual similarity (STS), an NLP technique for determining how similar two pieces of text are in terms of meaning, to analyze our \textit{ResearchBot's} performance and accuracy. The primary goal of this analysis is to compare the likeness in meaning between the user's answer to a question on SE Stack Exchange and \textit{ResearchBot's} generated response. The higher their similarity, the better our tool answers the user's question. To achieve our goal, we used the Sentence-BERT \footnote{\urlSBERT} (SBERT) model and the similarity method from spaCy \footnote{\urlSpacy}, an open-source NLP library in Python.

\subsubsection{Analysis Process}
We selected 100 question-and-answer sets for analysis. We first wrote a script to generate corresponding answer outputs for the 100 questions by \textit{ResearchBot}. Next, we used the STS tools, SBERT and spaCy, to compare the answers most upvoted from users (top answer) and answers produced by \textit{ResearchBot} (generated answer). We used the top answer to represent the correct answer, most accepted by humans with similar inquiries. Finally, we calculated each comparison set's semantic similarity score and each STS tool's overall statistic measurement characteristics of data. 
This analysis may introduce biases, as it is based on the assumption that the top answer is both correct and precise. This could lead to overestimating the bot's performance, as similarity with the top answer might not represent the most accurate or comprehensive response. To enhance the validity of our results, the incorporation of a broader range of benchmarks and metrics are recommended.

\subsubsection{LLM Comparison} 
To further assess \textit{ResearchBot}'s performance, we compared its responses with those generated by other LLM tools commonly used in SE: Google Gemini, GitHub Copilot, and ChatGPT. This evaluation involved the authors manually comparing the 50 top questions from Stack Exchange with responses from the selected LLMs. We specifically looked for the evidence provided to back up responses for each model. Upon manual review of the answers, we found responses from Gemini and Copilot were unreliable and lacked quality references. ChatGPT offered better resources; thus, as part of our future work we plan to conduct a user study in which SE practitioners evaluate responses from both \textit{ResearchBot} and ChatGPT. This will provide insights into user preferences regarding the clarity, reliability, and overall usefulness of responses, enabling us to refine \textit{ResearchBot} based on practitioner feedback and further tailor it to meet their needs.
\section{Results}
This section will present the results of our data collection and analysis approach. Table \ref{STSResults_StackExchnage} presents the statistical results for semantic similarity comparing the performance of \textit{ResearchBot} with Stack Exchange.

\begin{table*}
\centering
  \caption{Stack Exchange vs. ResearchBot}
  \label{STSResults_StackExchnage}
  \begin{tabular}{lccccc}
    \toprule
    \textbf{Semantic Textual Similarity (STS) Tools} & \textbf{Maximum} & \textbf{Minimum} & \textbf{Average} & \textbf{Median} & \textbf{Std Dev} \\
    \midrule
    {Sentence-BERT model} & {0.9147} & {0.0341} & {0.6609} & {0.6788} & {0.1618}\\
    {spaCy library} & {0.9473} & {0.7908} & {0.9203} & {0.9335} & {0.0506} \\
    \bottomrule
  \end{tabular}
\end{table*}

    
    

    
    
    

\subsection{STS Results}
The results presented in Table \ref{STSResults_StackExchnage} demonstrate that \textit{ResearchBot} performs effectively in answering users' questions as analyzed by spaCy, with an average similarity score of 0.92. However, the performance analyzed by SBERT shows a lower average of 0.66. This discrepancy is largely due to a notably low minimum similarity score of 0.0341 recorded by SBERT. This result highlights SBERT's higher scrutiny evaluation criteria, contributing to higher variability in scores, as indicated by a standard deviation of 0.16, compared to spaCy. Such variability suggests a re-evaluation of spaCy's data handling, as it indicates methodological differences, suggesting a need for a deeper comparative analysis of how each tool handles and interprets textual variations. Further analysis could provide insights into the different sensitivities of the tools to specific text features or contexts, thus explaining the variations in their performance metrics. Moreover, similar to recent efforts in sentiment analysis, future work can explore SE-specific NLP tools for STS.

\subsection{LLM Comparison}
The complete analysis for our manual analysis is available online.\footnote{\url{https://shorturl.at/QUAhn}} Overall, the answer outputs demonstrate \textit{ResearchBot} provides more reliable responses than other LLMs. For instance, we observed 12\% of responses provided by Gemini ($n = 6$) contained at least one invalid URL. Across GPT-4o, Gemini, and GitHub Copilot, we observed no references to academic resources in responses, with most relying only on blog posts and other non-credible sources~\cite{paperpile}---if evidence was provided at all. Meanwhile, all responses from \textit{ResearchBot} ($n = 50$, 100\%) incorporated scholarly literature in generated responses to a variety of SE-related questions. 

Our findings show \textit{ResearchBot} not only matches the top answers but also provides more comprehensive and detailed responses with research-based evidence. This makes our tool particularly valuable. Consequently, while \textit{ResearchBot} shows promise, there is room for improvements of our tool and evaluating its capabilities in answering SE-related questions.



\section{Discussion}
    \subsection{Implications of the Results}
    By utilizing STS tools and comparing against mainstream LLMs, we evaluated \textit{ResearchBot}'s ability to generate accurate and reliable responses. The results show mean STS scores of 0.66 (SBERT) and 0.92 (spaCy), indicating human-like responses to SE-related questions. Despite SBERT's lower scores, manual analysis confirms that \textit{ResearchBot} provides not only correct answers but also valuable background information. Our system's responses are comparable to ChatGPT while offering more reliable, research-backed answers. This study highlights the potential for evidence-based tools to enhance knowledge transfer between academia and industry in SE.

    \subsection{Future Work}
    We envision the following future work to further extend and evaluate \textit{ResearchBot}:

    \paragraph{User Study to Further Evaluate the Responses}
    A limitation of our preliminary study is we rely on STS comparisons with top answers from SE Stack Exchange and ChatGPT responses to evaluate the capabilities of \textit{ResearchBot}. As previously mentioned, we are currently conducting a user study where participants are asked to choose between answers to practical SE-related questions generated by \textit{ResearchBot} and ChatGPT, providing a rationale for their preference. This study aims to provide a better understanding of the types of answers users prefer, motivate enhancements to \textit{ResearchBot}, and further promote evidence-based practices in SE contexts. Future studies will include usability testing with software developers to assess the chatbot’s interface, response clarity, and practical relevance.

    \paragraph{Expansion of Academic Repositories}
    Currently, \textit{ResearchBot} only utilizes the Crossref repository to retrieving academic papers for summarization in responses to practical SE questions. Future iterations will incorporate additional repositories such as arXiv, ACM, IEEE, or academic journals in specific SE-related domains. This expansion would enrich the diversity of available research sources and broaden the scope of insights provided to users.

    \paragraph{User Interaction and Feedback Mechanisms}
    The preliminary prototype of \textit{ResearchBot} has limited features. Implementing interactive features, such as feedback loops, personalized recommendations, and the ability for follow-up inquiries and discussions in chat form similar to other LLM-based systems, can enhance \textit{ResearchBot}'s user experience. In addition, integrating mechanisms for users to rate responses and provide feedback will enable iterative improvements based on community input.


    \paragraph{Code Analysis}
    A limitation of our chatbot is that it does not have the ability to generate or understand code to provide responses to SE-related questions. Advancing \textit{ResearchBot}'s capabilities to understand, create, and/or correct code based on extracted insights from academic research can empower users to not only obtain theoretical knowledge but also practical code solutions aligned with research findings, enhancing evidence-based practices in industry.

    \paragraph{Integration with Programming Platforms}
    We also envision exploring integration opportunities for our system in popular programming platforms beyond SE Stack Exchange, such as GitHub or developer forums. This can expand \textit{ResearchBot}'s reach and utility within the broader SE community. Further, integration into integrated development environments (IDEs) can also facilitate insights from \textit{ResearchBot} during software development to answer questions and inform decision-making based on empirical research findings.

\section{Conclusion }
\textit{ResearchBot} applies advanced AI techniques to connect academic SE research with programming communities. By using NLP and intelligent summarization, our tool helps developers access and use relevant research insights more effectively. Our findings indicate that the early prototype of \textit{ResearchBot} provides accurate human-like responses to practical SE inquiries, comparable to ChatGPT with more reliable research-based evidence supporting responses. \textit{Research Bot} has the potential to transform how programmers interact with academic knowledge. The tool's ability to interpret user queries, search repositories, and summarize key findings from research papers can make scientific knowledge more accessible to practitioners, meeting immediate needs and helping tackle complex SE challenges with evidence-based responses.




\balance

\bibliographystyle{ieeetr}
\bibliography{main}

\begin{thebibliography}{10}

\bibitem{barua2014developers}
A.~Barua, S.~W. Thomas, and A.~E. Hassan, ``What are developers talking about? an analysis of topics and trends in stack overflow,'' {\em Empirical software engineering}, vol.~19, pp.~619--654, 2014.

\bibitem{sulir2022software}
M.~Sul{\'\i}r and M.~Regeci, ``Software engineers' questions and answers on stack exchange,'' in {\em 2022 IEEE 16th International Scientific Conference on Informatics (Informatics)}, pp.~304--310, IEEE, 2022.

\bibitem{menshikova2018evaluation}
A.~Menshikova, ``Evaluation of expertise in a virtual community of practice: The case of stack overflow,'' in {\em Digital Transformation and Global Society: Third International Conference, DTGS 2018, St. Petersburg, Russia, May 30--June 2, 2018, Revised Selected Papers, Part I 3}, pp.~483--491, Springer, 2018.

\bibitem{leung2023automated}
M.~Leung and G.~Murphy, ``On automated assistants for software development: The role of llms,'' in {\em 2023 38th IEEE/ACM International Conference on Automated Software Engineering (ASE)}, pp.~1737--1741, IEEE, 2023.

\bibitem{xiao2024devgpt}
T.~Xiao, C.~Treude, H.~Hata, and K.~Matsumoto, ``Devgpt: Studying developer-chatgpt conversations,'' in {\em 2024 IEEE/ACM 21st International Conference on Mining Software Repositories (MSR)}, pp.~227--230, IEEE, 2024.

\bibitem{kabir2023answers}
S.~Kabir, D.~N. Udo-Imeh, B.~Kou, and T.~Zhang, ``Who answers it better? an in-depth analysis of chatgpt and stack overflow answers to software engineering questions,'' {\em arXiv preprint arXiv:2308.02312}, 2023.

\bibitem{huang2024tale}
R.~Huang and S.~Chattopadhyay, ``A tale of two communities: Exploring academic references on stack overflow,'' in {\em Companion Proceedings of the ACM on Web Conference 2024}, pp.~855--858, 2024.

\bibitem{auer2023sciqa}
S.~Auer, D.~A. Barone, C.~Bartz, {\em et~al.}, ``The sciqa scientific question answering benchmark for scholarly knowledge,'' {\em Scientific Reports}, vol.~13, no.~1, p.~7240, 2023.

\bibitem{evidence}
T.~Dybå, B.~Kitchenham, and M.~Jørgensen, ``Evidence-based software engineering for practitioners,'' {\em Software, IEEE}, vol.~22, pp.~58 -- 65, 02 2005.

\bibitem{arcuri2011practical}
A.~Arcuri and L.~Briand, ``A practical guide for using statistical tests to assess randomized algorithms in software engineering,'' in {\em Proceedings of the 33rd international conference on software engineering}, pp.~1--10, 2011.

\bibitem{kitchenham2004evidence}
B.~A. Kitchenham, T.~Dyba, and M.~Jorgensen, ``Evidence-based software engineering,'' in {\em Proceedings. 26th International Conference on Software Engineering}, pp.~273--281, IEEE, 2004.

\bibitem{storey2020software}
M.-A. Storey, N.~A. Ernst, C.~Williams, and E.~Kalliamvakou, ``The who, what, how of software engineering research: a socio-technical framework,'' {\em Empirical Software Engineering}, vol.~25, pp.~4097--4129, 2020.

\bibitem{rainer2003persuading}
A.~Rainer, T.~Hall, and N.~Baddoo, ``Persuading developers to" buy into" software process improvement: a local opinion and empirical evidence,'' in {\em 2003 International Symposium on Empirical Software Engineering, 2003. ISESE 2003. Proceedings.}, pp.~326--335, IEEE, 2003.

\bibitem{devanbu2016belief}
P.~Devanbu, T.~Zimmermann, and C.~Bird, ``Belief \& evidence in empirical software engineering,'' in {\em Proceedings of the 38th international conference on software engineering}, pp.~108--119, 2016.

\bibitem{winters2024thoughts}
T.~Winters, ``Thoughts on applicability,'' {\em Journal of Systems and Software}, vol.~215, p.~112086, 2024.

\bibitem{lo2015practitioners}
D.~Lo, N.~Nagappan, and T.~Zimmermann, ``How practitioners perceive the relevance of software engineering research,'' in {\em Proceedings of the 2015 10th Joint Meeting on Foundations of Software Engineering}, pp.~415--425, 2015.

\bibitem{wilson2024will}
G.~Wilson, J.~Aranda, M.~Hoye, and B.~Johnson, ``It will never work in theory,'' {\em IEEE Software}, 2024.

\bibitem{access}
P.~Dunlop, ``How to access academic research when you’re not affiliated with a university,'' {\em Linkedin Pulse}, 2019.
\newblock \url{https://www.linkedin.com/pulse/how-access-academic-research-when-youre-affiliated-patrick-dunlop/}.

\bibitem{horvitznow}
E.~Horvitz, V.~Conitzer, S.~McIlraith, and P.~Stone, ``Now, later, and lasting: 10 priorities for ai research, policy, and practice,'' {\em Communications of the ACM}.

\bibitem{elastic}
``What are large language models (llms)?,'' {\em Elastic}, 2024.
\newblock \url{https://www.elastic.co/what-is/large-language-models}.

\bibitem{paperpile}
Paperpile, ``Credible sources: what are they and how to identify them,'' {\em Paperpile Research Guide}, 2024.
\newblock \url{https://paperpile.com/g/what-are-credible-sources/}.

\end{thebibliography}


\end{document}